# Databases for comparative syntactic research

Jessica K. Ivani Balthasar Bickel

University of Zurich

## Abstract

Recent years have witnessed a steep increase in linguistic databases capturing syntactic variation. We survey and describe 21 publicly available morpho-syntactic databases, focusing on such properties as data structure, user interface, documentation, formats, and overall user friendliness. We demonstrate that all the surveyed databases can be fruitfully categorized along two dimensions: units of description and the design principle. Units of description refer to the type of the data the database represents (languages, constructions, or expressions). The design principles capture the internal logic of the database. We identify three primary design principles, which vary in their descriptive power, granularity, and complexity: monocategorization, multicategorization, and structural decomposition. We describe how these design principles are implemented in concrete databases and discuss their advantages and limitations. Finally, we outline essential desiderata for future modern databases in linguistics.

## Introduction

Recent years have witnessed a steep increase in linguistic databases capturing syntactic variation. The goal of this contribution is firstly to categorize the approaches in this endeavor, focusing on both design choices and practical aspects, and secondly to offer a survey of a selection of publicly available online databases.

The chapter is structured as follows: we first discuss and typologize the recurrent design principles we recognise in databases of linguistic variation (Section 1). The discussion of other criteria used in our survey, such as user interface, documentation and data formats, is provided in Section 2. The survey itself, with resources organized thematically, follows in Section 3. We conclude the contribution with a summary and desiderata for future modern databases (Section 4). A summary table of the surveyed databases and their properties is provided in the Appendix.

## 1. Database design principles and units of description

A database is a system that stores descriptive statements representing knowledge about some domain of interest. One of the primary functions of a database is to allow quick data search and retrieval according to criteria or patterns of interest, which requires the stored information to be appropriately structured and systematically organized. There are different ways to achieve this, depending on the nature of the data in question and the overall goal. In other words, designing a database is about making choices about what is being represented and the mechanisms by which it is represented.

Surveying current databases on syntax suggests that they vary in two key dimensions: units of description and design approaches, with nine basic combinations. Each of these combinations is amply attested in the databases we survey in this chapter (Table 1).

| | | Design | | |
|---|---|---|---|---|
| | | **Monocategorization** | **Multicategorization** | **Structural decomposition** |
| Units | **Language** | WALS, Chapter 33: Coding of nominal plurality | Grambank: feature subset on nominal plurality | AUTOTYP Verb Synthesis module |
| | **Construction** | WALS, Chapters 125-129: Subordinate clauses | SAILS Subordination (SUB) domain | Tymber: grammatical number |
| | **Expression** | ASIt: types of adverbs | Afranaph: Clausal Complementation | DAI: agreement |

*Table 1: Units of description and design approaches combinations as exemplified in existing linguistic databases*

**Units**

What is being represented is usually some aspect of an individual language as whole, e.g. the amount of morphology in English, of an individual construction in a language (however it may be captured, e.g. by the rules that generate it, or the constraints it is subject to), e.g. the presence of number agreement in finite main clauses), or a concrete linguistic expression (token), e.g. *the paper reads well*. We call these the "units of description", and call "item of description" any element belonging to these units. In our survey, we identify languages, constructions and expressions as units of description. Most databases choose to focus on one of these units, while others can be heterogeneous.

Items in a database are described by associating them with a system of variables (also known as 'parameters', 'features', or 'categories'). These associations can take a number of forms: type assignments (using a finite, predefined set of types), binary characterizing statements (e.g. *does x have property y* or *can x be described as y*), numeric data (such as counts and ratios), identifiers (to identify unique items or systems), and relational references that link together different items (or groups of items). For example, annotating an expression through a Part-of-Speech tag is a type assignment, where the Part-of-Speech tag is part of a predefined set of types.

**Designs**

In the databases in our survey, we identify three basic design principles, which we refer to as monocategorization, multicategorization and structural decomposition.[1] The purpose of this classification is to aid understanding the rationale and logic behind the individual databases. In other words, the distinctions we make represent idealized conceptual scenarios, often followed unconsciously by the database compilers. Real-world databases are often fluid in their design and may exhibit properties of more than one of these designs. Thus, when describing the design principle of each database, we focus on the approach that best captures the overall logic of that database.

A **monocategorizing** database maps linguistic phenomena onto a set of concrete types (categories). The basic idea is to assign an item to exactly one type, hence 'monocategorization'. This is schematically illustrated in Figure 1. A monocategorizing database can still have multiple variables, each describing a specific aspect or domain; however, each variable involves an unique assignment of a type to an item and the database includes no statement about how variables relate to each other. This approach aims to capture the broad essence of a phenomenon or domain, often trading details for simplicity.

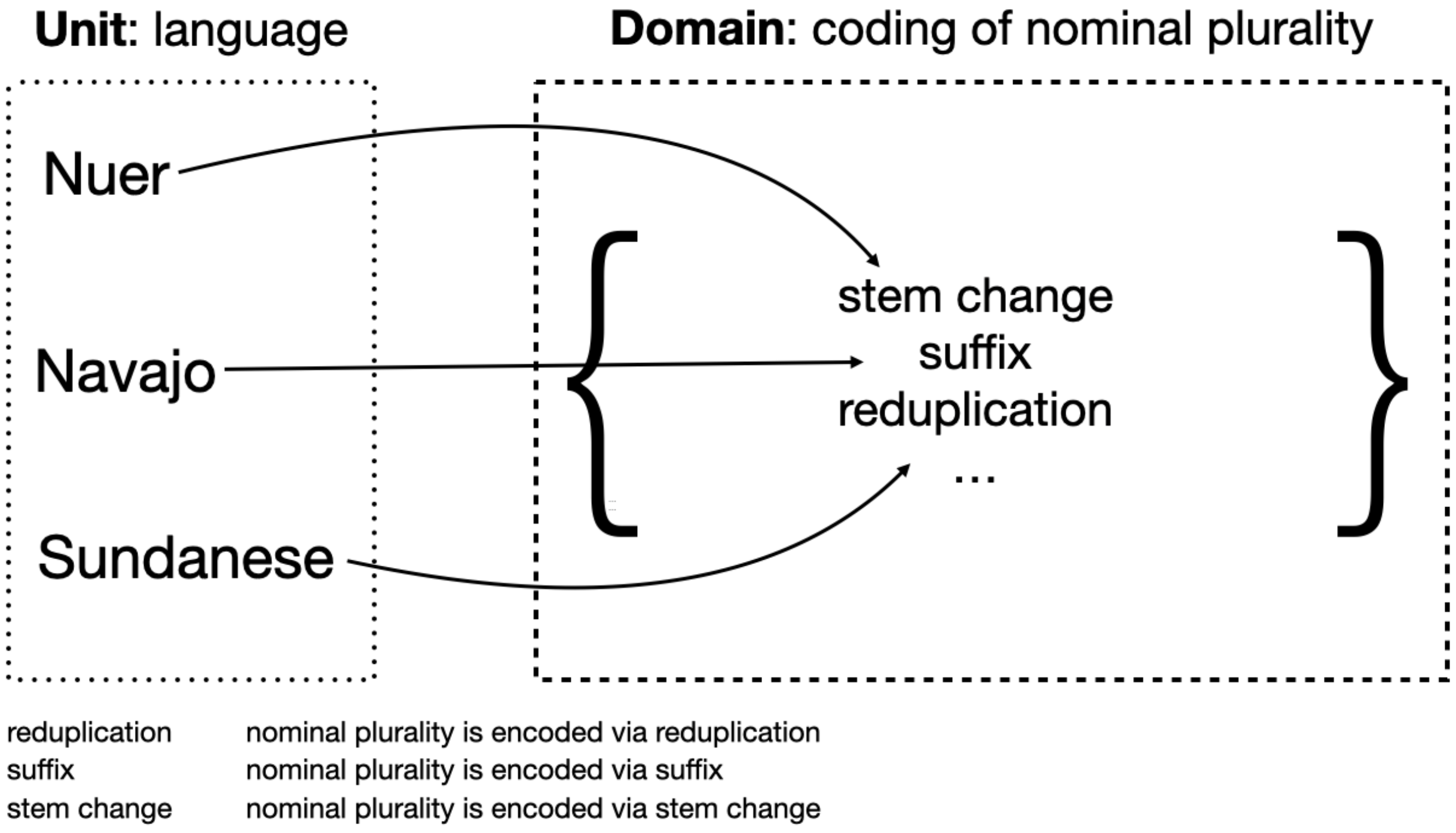

*Figure 1. Monocategorized representation of Units → Domain in WALS 33 Chapter "Coding of nominal plurality"*

[1] It should be noted that this database classification only pertains to the datasets surveyed in this contribution and does not represent all the possible approaches to conceptualizing databases. For example, in the early 2000s there was a short-lived drive to design databases as ontologies using the Web Ontology Language (OWL, see Farrar & Langedoen 2003; Cysouw et al. 2005). Since that particular approach was abandoned fairly quickly and the relevant databases are not retrievable or useable without significant archeaological effort, we will not cover it here.

Many chapters of The World Atlas of Languages Structures (WALS, Dryer & Haspelmath 2013) rely on monocategorization to describe linguistic features. Nominal plurality in WALS (Chapter 33, Dryer 2013) is partitioned into nine types that correspond to various means of coding plurality, such as suffix, prefix, stem change, and the like. The unit of description, in this case the language, is assigned to one type. For example, Navajo (Na-Dene, Glottocode [nava1243] is assigned the type suffix, which encodes the descriptive fact that nominal plural in Navajo is expressed by suffix. The monocategorization approach is not exclusive to languages. An example of monocategorization for constructions is found again in WALS Chapter 125, which describes purpose clauses (Cristofaro 2013). Purpose clauses are partitioned into three types: balanced, deranked and balanced/deranked. Each construction is assigned a type. For example, Abhkaz (Northwest Caucasian, Glottocode [abkh1244]) purpose clauses are assigned the type balanced/deranked.

Finally, Atlante Sintattico d'Italia (ASIt, Poletto and Benincà 2007) is an instance of monocategorization applied to linguistic expressions. In ASIt, adverbs are partitioned into types (aspectual, locative, modal, and other). Each expression is assigned a type. The expression "*Gli devo parlare subito*" ("I need to talk to him/them immediately") is assigned the type 'temporal adverbs'.

**Multicategorization** is a straightforward extension of monocategorization which retains simplicity while improving descriptive power. Rather than partitioning the domain space into distinct types, multicategorization uses multiple such partitions along different dimensions. In practice, these dimensions are often characterizing statements, such as *does x have the property y?* or *can x be described as y?*. An item is assigned a type from each of these dimensions and its description exists as the intersection of the individual categories. Because of this, the relationship between variables is explicitly encoded in the database. Where monocategorization reduces a domain to a single type, multicategorization reduces the domain to an intersection of multiple (overlapping) categories, hence the name. Grambank (Grambank Consortium 2022) uses multicategorization to represent linguistic phenomena, with languages as units of description. The Grambank design is schematically illustrated in Figure 2.

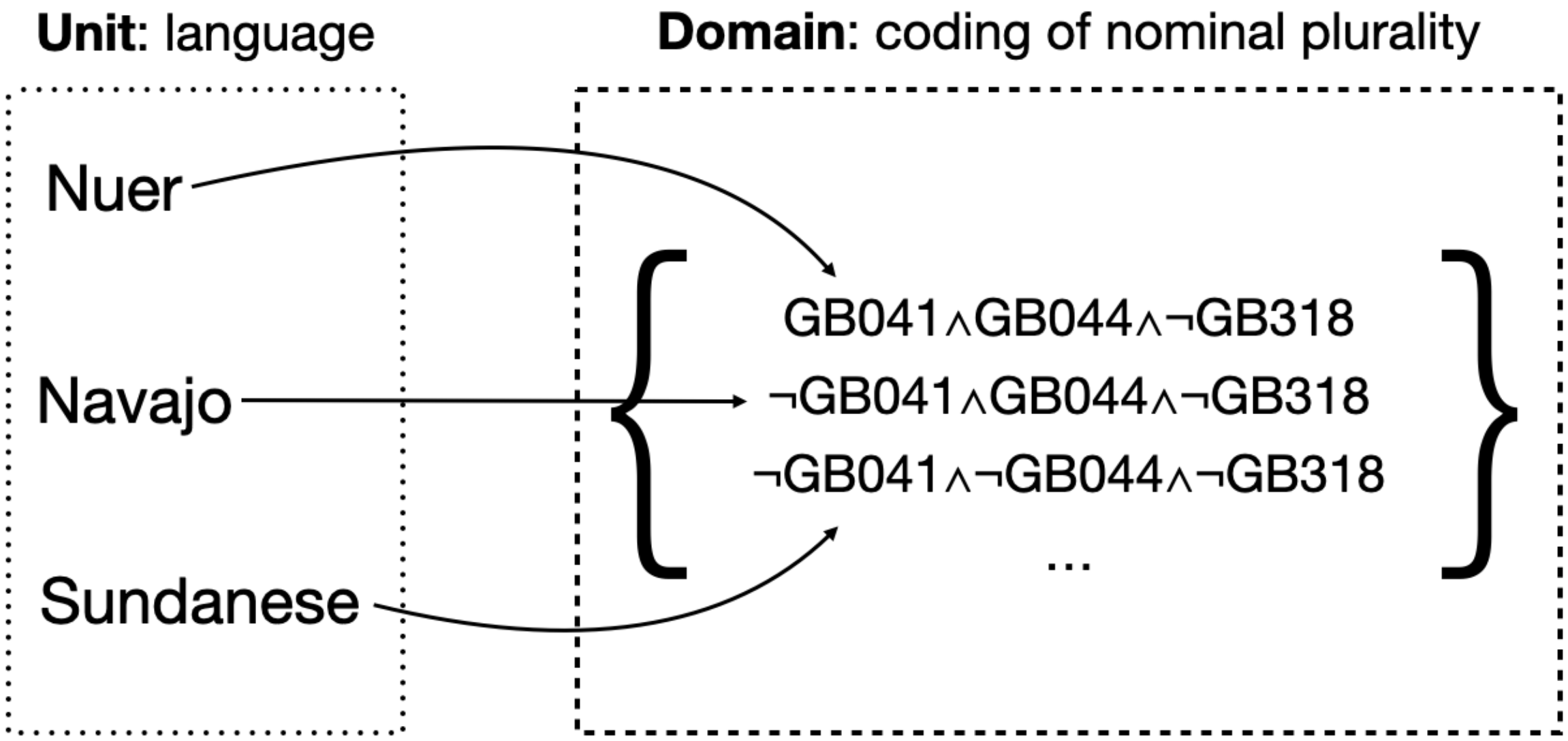

| | |
|---|---|
| GB041 | Are there several nouns (more than three) which are suppletive for number? |
| GB044 | Is there productive morphological plural marking on nouns? |
| GB318 | Is plural number regularly marked in the noun phrase by a dedicated phonologically free element? |

*Figure 2. Multicategorized representation of Units → Domain in Grambank*

Grambank uses multicategorization to represent linguistic phenomena, with languages as units of description. Categories are primarily described by binary characterizing statements (identified by feature IDs). As exemplified above, nominal plurality in Grambank is partitioned into overlapping categories, represented by the answers to the following questions:[2] GB041: *Are there several nouns (more than three) which are suppletive for numbe*r?; GB044: *Is there productive morphological plural marking on nouns*?; GB318: *Is plural number regularly marked in the noun phrase by a dedicated phonologically free element*?. The units, here languages, are assigned a value from each of these categories, and the final description is the intersection thereof, thereby explicitly linking each category with each other. For example, Navajo has value "yes" (1) for GB044 and "no" (0) for GB041 and GB318. The resulting description is then "Najavo has productive plural marking on nouns, does not have several nouns that are suppletive for number and does not mark number by a phonologically free element".

The same approach can be applied to constructions and expressions. An example of the former is the subordination dataset (SUB) within the South American Indigenous Language Structures (SAILS) database (Muysken et al. 2016). In SUB, the subordination domain is partitioned into a rich set of overlapping categories. Each construction is assigned the intersection of these categories that make up the description.

Afranaph (Safir 2008) applies the multicategorization approach to expressions. In Afranaph, phenomena such as complement clauses and anaphora are partitioned into several categories, defined through binary and categorical statements. Expressions are assigned the corresponding overlapping categories.

[2] This list of statements is not exhaustive, as other features describing noun plurality are included in Grambank. For convenience, we restrict our exemplification to the Grambank features GB041, GB044, and GB318.

Whereas monocategorization and multicategorization manage complexity by assigning phenomena to types, a fundamentally different approach is to break down (decompose) the phenomenon into its intrinsic structural parts and describe these individually. For example, suppose that we want to describe paradigm structures. Using mono- or multi- categorization principles we could characterize a paradigm by the presence or the counts of specific features (e.g. *Any prefixes? Any prefixes for number? How many prefixes?* etc.). The **structural decomposition** approach foregoes these assignments and instead describes the properties of the paradigm's cell fillers independently of each other (e.g. Is third person singular past tense expressed by a prefix? Is third person plural past tense expressed by a prefix? etc). Critically, however, the database furthermore codes that these cells belong to the same paradigm, allowing derived descriptions (e.g. counting the proportion of cells with prefixal forms in each paradigm). What we obtain this way is a structural decomposition, a representation of the phenomenon that is close to how one would describe the phenomenon in a reference grammar or a formal syntactic analysis. The result is typically a series of nested descriptions, e.g. a cell in a paradigm might contain descriptions of each category in terms of their semantics or form. This approach is known as multivariate typology (Bickel and Nichols 2002), but structural decomposition captures better the difference from multicategorization.

A typical database of this kind is the AUTOTYP Verb Synthesis module (Bickel et al. 2021), illustrated in Figure 3, with Belhare (Sino-Tibetan, Glottocode [belh1239]) as a concrete example.

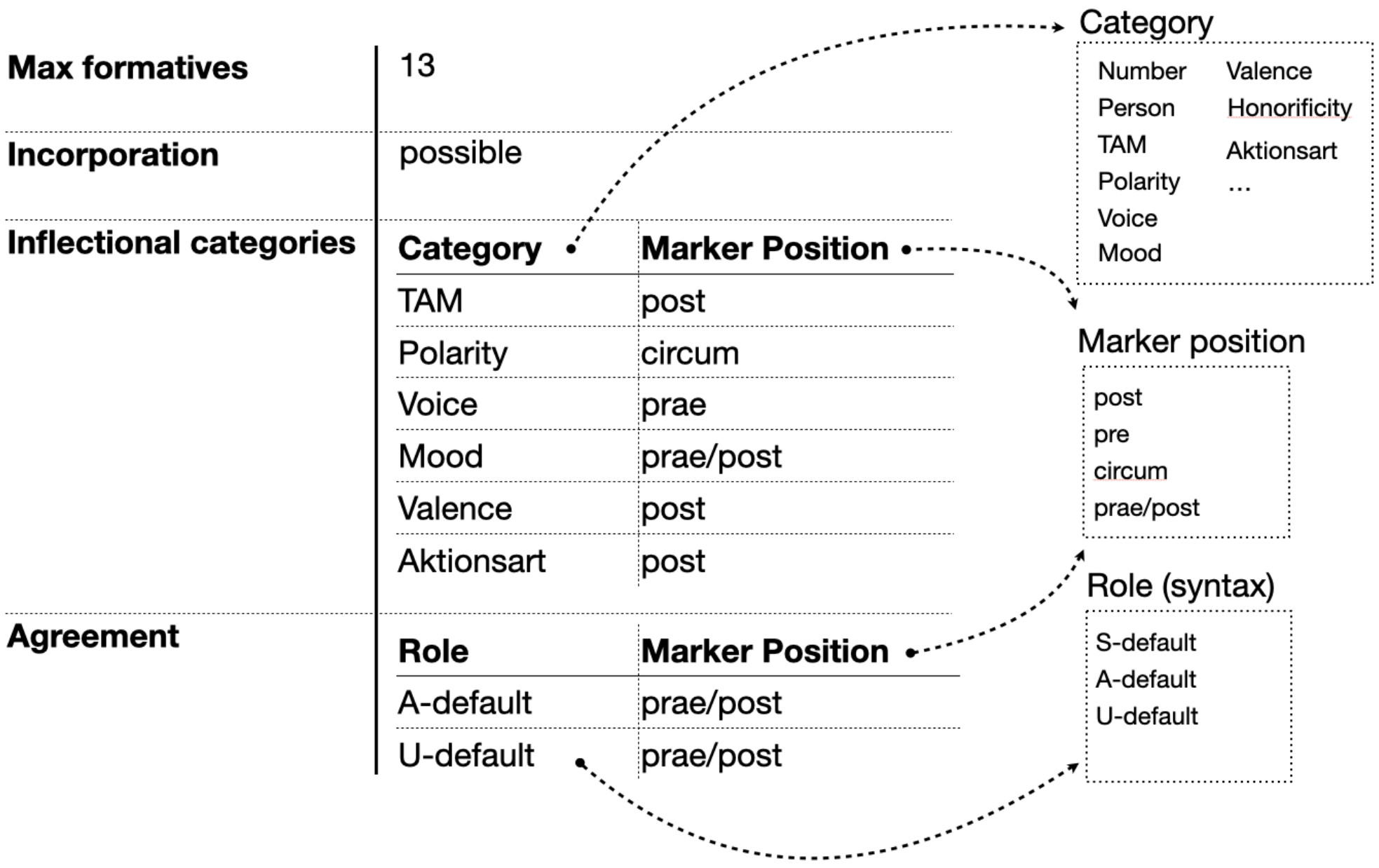

Figure 3. Structurally decomposed representation of *Units → Domain in AUTOTYP Synthesis Module*

In the AUTOTYP Verb Synthesis Module, the verb domain is broken down ("decomposed") into structural properties, here corresponding to Categories, Marker Positions and Roles. These structural properties are described in a composite, nested fashion and are linked to each other through relational graphs. The derived complex schema makes up the domain and is here assigned to the unit of description, here the language, where summary statements can be derived (e.g. on the count of categories, or the count of post-positioned role markers)
This design is also followed by Tymber (Ivani & Zakharko 2019) with respect to constructions. Tymber uses an interlinked relational schema to combine together descriptions of individual markers, number marking contexts, and classes of nominals in order to provide a fine-grained description of nominal number marking.
Finally, we find the same approach applied to linguistic expressions in the Zürich Database of Italo-Romance (DAI, Loporcaro et al. 2018). The agreement relations in DAI are represented using graph structures that link together different linguistic expressions and their properties.

Different choices of the unit of description and the design approach represent various trade-offs between simplicity and descriptive power. These also have important consequences for the databases' user-friendliness and the long-term reusability of the data. Databases that pair languages with a monocategorizing approach are the simplest and usually among the most intuitive. Most of these databases include a ready-to-use web interface for data exploration and are usually easy to grasp to novice users. At the same time, these databases often suffer from oversimplification. By reducing rich phenomena to just a few types, some complexity is lost, and users have no way to capture what falls out of the predefined types, either by way of internal variation or by missing rare types. Also, it remains unclear how variables relate to each other. They might be logically independent or not (e.g. a variable on properties of case markers depends on a variable registering the presence of case), but this is not explicitly represented in the database. There is a risk that such dependencies go unnoticed when performing analyses across many variables together.

Databases that rely on multicategorization or structural decomposition can capture phenomena at a higher resolution. These resources, being less constrained to pre-aggregated types, offer broader opportunities for data exploration and analyses, allowing the users to get insights into phenomena that go beyond the original purpose of the database. Crucially, data aggregation is only possible from a more involved ("fine") data model to a less involved one. By collecting granular data and aggregating it at a later stage, it is often possible to manipulate and compress the data into more specific, simplified types as required by the purpose at hand. This property makes databases that collect fine-grained data more powerful and easier to reuse, improving their long-term value. Conversely, the more granular the description, the more involved the database structure and interface become. This is especially true for databases relying on structural decomposition. These resources usually represent more complex and less intuitive variables that require more involved tools and technical skills for their exploration and analyses. In addition, complex variables require comprehensive documentation to support users in exploring the database.

To some extent the design approaches discussed here bear some relationship to the spirit of various syntactic theories. For example, the Principles and Parameters framework loosely corresponds to the multicategorization approach, where a grammar is represented as a (structured) conjunction of parameters. Approaches like HPSG, by contrast, are more comparable to structural decomposition, where a grammar is represented as attribute value matrices linked with each other in specific ways. These analogies shouldn't be over-interpreted, however, and it is clear that statements in a databse can be transformed into the analytical statements made by syntactic theories (though of course with varying degrees of precision and ease).

# 2. Coverage, topics, documentation and data export

**Coverage and topics**

Databases vary in topics covered: some may deal with an in-depth overview of a syntactic phenomenon or a set of related linguistic phenomena, while other resources aim at covering a wide range of linguistic domains. Examples of feature-specific databases include ValPal (Hartmann et al. 2013) on valency, and Tymber (Ivani and Zakharko 2019), targeting grammatical number. Databases covering several domains include AUTOTYP (Bickel et al. 2021), Grambank (Grambank Consortium 2022), and WALS (Dryer and Haspelmath 2013).

In addition, databases vary in genealogical and geographic focus. Our survey includes language-specific databases exploring micro-syntactic complexity and diatopic and diastratic variation. Databases of this kind are the BiV database for Basque (Orbegozo et al. 2018) and DynaSAND for Dutch (Barbiers et al. 2006), among others. Several existing databases investigate diversity within a single language family, for example, the UraTyp project (Uralic language family; Norvik et al. 2022). Other resources focus on geographical areas characterized by genealogically diverse languages, such as Afranaph (describing the languages of Africa, Sapir 2008). Finally, we illustrate databases as part of large-scale typological projects (AUTOTYP, WALS, and Grambank, among others).

**Interface, operability, documentation, and data export**

We categorize more technical aspects of databases, such as the presence of documentation, query interfaces, and their usability.

We report the exhaustiveness of the documentation and the metadata that support each database. We illustrate the informativeness of the documentation, the level of detail, and how it helps and guides the user in understanding the database and its variables. We distinguish between databases with minimal and limited documentation (where the understanding of the database structure and rationale is left to the users' efforts) and those resources with detailed documentation (the database structure is explained in detail and the variables are clearly defined and documented).

Online databases are often accessible through a query interface, while others offer downloadable repositories of datasets explorable by scripting data processing steps, e.g. in R or Python. Several databases provide both solutions, allowing users to analyze the data online and offline. We describe the query interfaces available, from essential hyperlink solutions to search functions, including predefined keywords, filtered and free searches, and complex queries. We also emphasize the availability of cartographic options.

We evaluate the usability of the query interface. We distinguish between easy and agile interfaces requiring little or no training and involved interfaces, which are more complex to navigate and not immediately graspable by the novice user.

A core feature of some databases is the offline access to the data that make up the datasets. We distinguish between resources that offer this option and those that restrict data exploration to the respective online interface. We illustrate the raw data formats available to the users: from CSV worksheets to JSON and R workspaces. The development of standardized data formats across databases and a unified experience in the query interface is one of the main goals of the Cross-Linguistic Linked Data (CLLD, Forkel et al. 2020) project, achieved by the development of the Cross-Linguistic Data Format framework (CLDF; Forkel et al. 2018). We list the databases following these guidelines and the solutions provided by non-CLDF-compliant databases. We also describe whether databases provide access to the raw data that make up the features values, such as language reports and example sheets.

Finally, we survey the ability to use data from different databases together. We survey the possibility of linking datasets from various databases for further exploration and analysis. For example, connectivity is achievable by using standardized IDs, such as ISO 693-3 codes and Glottocodes from Glottolog (Hammarstrom et al. 2019).

# 3. Survey of syntactic databases

In what follows, we provide a survey of freely accessible online databases containing enough data to perform substantial comparative analysis. We exclude from our survey scattered datasets disseminated in research papers, and we focus on publicly available and operational databases. We present the databases thematically. We first describe language-specific databases (Section 3.1), areal databases on an individual language family or dialectal continuum (Section 3.2), and language family-specific databases (Section 3.3). We move to broader databases that include data from genealogically diverse languages spoken in a wide area (Section 3.4) and typological databases, both feature-specific and covering an extensive range of morphosyntactic domains (Section 3.5). A summary Table with a list of the surveyed databases and their main properties described along the criteria outlined above can be found in the Appendix.

## 3.1 Language-specific databases

Several resources deal with the syntactic micro-variation within a language. These dialectal databases focus on diatopic variation and explore the structural morphosyntactic deviations from the standard language. Databases of this kind include the Basque in Variation database, the DynaSAND project for Dutch, SADS for Swiss German, and the eWAVE 3.0 project for English.

### Basque in Variation (BiV)

The Basque in Variation (BiV; https://basqueandbeyond.ehu.eus/biv/; Orbegozo et al., 2018) database explores the syntactic (micro) variation of Basque, by collecting more than a hundred morphosyntactic features in several varieties of Basque. The varieties are linked in the database to a unique internal identifier (the BiV code), a geographical data point, and the dialect area. The data comes from a questionnaire structured around grammaticality judgments values (in binary yes-no form) to a list of sentences representing a morphosyntactic variable. The morphosyntactic variables described in BiV are grouped into broader domains (corresponding to the Chapters in the online BiV interface). Domains include case and agreement, auxiliary alternation, aspect, postpositions, transitivity, complementizer, and mood. Variables ("Features" in BiV) are described by distinguishing a `generalized` pattern, corresponding to the predominant variant attested in standard Basque, and the `variational` pattern(s), which collects the attested variation(s) from the standard. Text examples describe both types. Pattern distributions are explorable in an interactive map. The Answer tab collects the data by variable: it contains the BiV code, geographic information, and the binary yes-no value. The description of the variational pattern(s) does not aim to be granular: it rather "corresponds to the most basic form of that feature, that is, the minimal form shared by all the varieties exhibiting the pattern" (Fernandez et al. 2019: 354).
Users can search data and feature(s) of interest via the web interface by predefined keywords that come from standard linguistic terminology (`addressee` or `absolutive`). The interface does not allow for simple or complex field queries. The binary values, organized in individual tables for each feature, are exportable in several formats (CSV, XLS, XLM). A full overview of the questionnaire by variety, complete with examples, is available in PDF.

### DynaSAND

The Dynamic Syntactic Atlas of the Dutch dialects (DynaSAND; https://www.meertens.knaw.nl/sand; Barbiers et al. 2006) is an online tool developed for the exploration of syntactic variation in the varieties of Dutch spoken in the Netherlands, Belgium, and north west France.
The DynaSAND web interface consists of a database, a search engine, a cartographic component, and a bibliography. In DynaSAND, language varieties are identified by a Kloeke-code, corresponding to standardized indicators of geographical places used in Dutch dialectology. The data collected in DynaSAND comes from several sources. These include written postal surveys, oral interviews at the locations of the dialects, and telephone interviews, organized at different stages, each with their respective questionnaires and informants pools. The questionnaire comprises translation tasks, indirect grammaticality judgments, meaning questions, and picture response tasks (uniquely identified in the online database, Barbiers et al. 2007:61). The data collection involved hundreds of informants distributed across hundreds of locations. For the oral and telephone interviews, the network of informants fulfills sociolinguistic criteria, such as age range and education level. The syntactic topics explored in DynaSAND cover several empirical domains and subdomains: the left and right peripheries of the clause, the (morpho-)syntax of verbs, negation, quantification, pronouns, wh-clauses, relative clauses (Barbiers et al. 2007:37).
The navigation engine of the DynaSAND database allows searches within each questionnaire (the written, oral, and telephone questionnaires) by querying Kloeke-codes, sentence codes, expressions (sentences in DynaSAND), and strings of PoS-tags. Users can filter searches by geographical municipalities and metadata. Moreover, each questionnaire is searchable through queries specific to the selected questionnaire's elicitation task, such as 'fill in the blanks' questions for the written questionnaire or expression tags and lemmas in the oral questionnaire. In addition, users can query for syntactic phenomena (predefined through text strings). Queries results show expressions organized by Kloeke-code, speaker information, transcriptions, context, and the audio recording of the elicited sentence. Additional information may include PoS tagging for each elicitation. DynaSAND allows the user to draw detailed customizable maps and compare the geographical distributions of features.

The DynaSAND database is integrated into MIMORE (https://www.meertens.knaw.nl/mimore/; Barbiers et al. 2016). MIMORE search engine contains a PoS tag constructor and predefined complex PoS tags to make searches with strings of tags possible. Through MIMORE, users can visualize the search results through tables or maps and export the result queries in CSV and XLS for offline use.

### Syntaktischer Atlas der deutschen Schweiz online (SADS online)

SADS online (https://dialektsyntax.linguistik.uzh.ch; Seiler et al. 2022) is an interactive atlas based on the Syntaktischer Atlas der deutschen Schweiz (SADS; Glaser 2021) describing the diatopic syntactic variation of Swiss German dialects. SADS contains data surveys from thousands of informants distributed over three hundred locations in the Swiss-German-speaking area. SADS covers several syntactic domains: the nominal phrase (including possessive constructions), case and pronouns, the verb domain, secondary predication, and simple and complex clauses. The data collection relies on more than a hundred different questions distributed over four written postal questionnaires, supported in some cases by oral materials. The questionnaires consist of translation tasks, variant choice and grammaticality, and ‘naturalness’ judgments of predefined Swiss German expressions. The interactive atlas plots on maps the values for each geographical data point, organized by questionnaire entries. In the online interface, users can browse the questions (uniquely identified and distinguished between translation and choice task) and plot the geographical distribution of the corresponding values. Each question is supported by a detailed description of the content of the questionnaire assignment and related keywords. Users can perform complex queries and plot values of combined questions. SADS data can be explored through the web interface and, for individual questions, the basic data displayed on the map is downloadable in CSV format for offline use. The maps created by the users can be exported as PDF files. The questionnaires are also available in PDF.

### eWAVE

The Electronic World Atlas of Varieties of English (eWAVE 3.0; available through a CLLD web application at https://ewave-atlas.org; Kortmann et al. 2020a) is an interactive atlas on morphosyntactic variation in spoken varieties of English worldwide. eWAVE is updated in irregular intervals; currently, it comprises almost a hundred varieties of English in eight Anglophone world regions (Africa, Asia, Australia, British Isles, Caribbean, North America, Pacific, and the South Atlantic). The languages in eWAVE are grouped into five broad Types: Traditional L1 varieties (including regional varieties spoken in Scotland, Ireland, and the British enclaves), High-contact L1 varieties (such as colloquial American English), Indigenized L2 varieties (e.g. Jamaican English), and English-based Pidgins (e.g. Tok Pisin) and Creoles (e.g. Bislama). For each variety, expert-based frequency ratings on the presence and pervasiveness of several morphosyntactic features from 12 grammatical domains are provided. This includes, for instance, information on discourse and word order, tense and aspect, adverbial subordination, and complementation. In addition, a short description and, if available, a Glottocode for the variety is provided. eWAVE was compiled based on surveys by expert contributors. The interface can be explored by items (with languages as the main units of description, variables ("Features" in eWAVE), informants, examples, and sources. Users can map the distribution of a specific variable across languages or by world region, and can plot value-feature combinations across the sample. The aggregated data is downloadable in CLDF StructureDataset (Kortmann et al. 2020b) for offline use, and the questionnaires are accessible to the users in PDF.

## 3.2 Language-continuum databases

Several databases narrow their focus to a specific area or a closely related group of languages. Resources of this kind include Atlante Sintattico d'Italia, the Scandinavian Dialect Syntax database, and the Zurich Database of Agreement in Italo-Romance.

### ASIt

Atlante Sintattico d'Italia (ASIt; http://asit.maldura.unipd.it) is a project developed by several research teams whose primary goal is to describe the syntactic variation of the languages spoken in Italy. The project, launched initially as ASIs (Benincà and Poletto 2007), with a focus on the languages of northern Italy, was later expanded to include hundreds of localities (and the corresponding varieties) in the Italian peninsula. The data consists of several questionnaires developed at different stages and spoken interviews. Some of these questionnaires explore generic syntactic domains such as relative clauses, nominal adverbial, prepositional syntax, exclamatives, imperatives, and negation. Other questionnaires target domains specific to some areas or dialects (i.e., subject clitics and prepositional objects). The questionnaires are structured as a list of standard (or regional) Italian expressions and are available on the ASIt web page.
In the ASIt database, expressions append to linguistic keywords (in Italian) that describe the underlying generic syntactic phenomena used for search and query purposes. The ASIt search tool allows users to filter through the entries by region and narrow queries to a specific province and municipality. Similarly, it is possible to restrict the research to linguistic genealogical subgroups (such as Gallo-Italic or Italo-Romance). ASIt allows the users to choose and combine several keywords or strings that describe the phenomena of choice: keywords include generic strings such as 'adverbs,' or 'quantifiers,' among others. Each of these keywords opens more restricted options (such as different types of adverbial constructions or negation structures). The database interface allows for complex and/or queries across feature(s) groups.
The search query lists results by expressions, by regional language, each with the respective standard Italian translation. The expressions in Italian dialects are transcribed orthographically. Results include details about the dialect where the phenomenon is attested, the municipality, and the linguistic region. The data is not exportable, and the user can explore the results exclusively through the website interface, available only in Italian.

### ScanDiaSyn

The Scandinavian Dialect Syntax (ScanDiaSyn; http://websim.arkivert.uit.no/scandiasyn/, accessible through CLARIN) is a broad-scope project with the primary goal being the systematic exploration of the syntactic variation across the Scandinavian linguistic continuum. ScanDiaSyn stems from the collaboration of several research groups in Scandinavian language studies. One of the project's outcomes, consultable on the ScanDiaSyn webpage, is the Nordic Syntax Database (Lindstad et al. 2009).

The Nordic Syntax Database contains morphosyntactic data on the diatopic variation of Danish, Faroese, Icelandic, Norwegian, and Swedish languages. The data comes from different sources and methodologies, from speakers' intuitions and grammaticality judgments to transcribed audio and video recordings, extracted from interviews, conversations, and translations of expressions from the standard language to the local variety of the informant. The sociolinguistic pool of the informants and the amount of data collected varies according to the language. ScanDiaSyn has gathered data at about 300 measure points in Scandinavia, with the average number ranging from 140 to 240 expressions tested at each data point (Lindstad et al. 2009: 283). In the Norwegian corpus, information about the speakers (such as sex and age range) is integrated into the data, allowing to control for diastratic and sociolinguistic properties.

The test expressions are inspired by a generative syntax approach based on speakers' evaluations. These cover several syntactic domains: binding and coreference, left and right periphery, subject placement, object shift, noun and verb phrase, and verb placement, among others.

Data is searchable by syntactic domain, flagged by keywords or text strings appended to the expressions (such as 'reflexives,' or `binding`). The search tool allows users to narrow queries to a specific language and locality (indicated by town). Additional filters include a 1-5 score that describes the value reported by the informant on the `acceptability scale` and the possibility to filter data by sociolinguistic variables. The expressions in the database, uniquely defined, come with an English translation and a transcription in the available Nordic languages. The lemmatization and PoS tagging are not consistent across the expressions of the database and vary depending on each language subproject. The query results can be visualized internally on the website interface and can not be exported or downloaded for offline use.

### Zurich Database of Agreement in Italo-Romance (DAI)

The Zurich Database of Agreement in Italo-Romance (DAI; https://www.dai.uzh.ch/new/#/public/home; Loporcaro et al. 2018) collects and describes agreement phenomena at a fine-grained level in a sample of Italo-Romance languages spoken in central and southern Italy. The choice of these languages is motivated by the peculiar and unique agreement patterns they display (Idone 2018). The DAI database follows the theoretical and terminological framework in agreement patterns proposed in Corbett (2006).

The database contains hundreds of morpho-syntactically annotated files (via ad hoc software) and >100,000 expressions (“tokens” in DAI), based on hundreds of hours of elicited and semi-spontaneous speech collected through questionnaires and picture stories.

The DAI interface allows users to search the data by agreement pattern or token query. In both environments, users can filter the data by datapoint(s), speaker, and source (questionnaire or semi-spontaneous speech).

Through the agreement query, users can narrow searches to the syntactic domain(s) or syntactic configuration(s) in which the agreement pattern occurs. These syntactic domains are a closed set of configurations defined by keywords (i.e., "infinitive agreement," "adverbial agreement"). In addition, the user can specify the morphosyntactic properties of the controller and the target. These properties include PoS, grammatical relation, position to the verb, and several others, depending on the properties of the PoS/Phrase selected as controller/target.

The token query enables advanced queries for any annotated expression in the dataset, searchable by gloss or form. Users can search for a specific lexical type glossed in the database in the gloss field. The DAI website contains documentation materials. These include individual overviews of the Italo-Romance varieties covered in the database (downloadable in PDF), transcription conventions, and sociolinguistic metadata. The search queries and the metadata results are not downloadable or exportable and are accessible through the website interface only. Audio data is available on request.

## 3.3 Family-specific databases

### UraTyp

The Uralic Areal Typology Online (UraTyp; available through a CLLD web application at https://uralic.clld.org; Norvik et al. 2022a) is a family-specific database exploring the linguistic variation of several Uralic languages through hundreds of structural features. The data comes from descriptive grammars and other reference materials. The variables ("Features" in UraTyp) list stems from two different questionnaires. One is the Grambank (see below) questionnaire (GB in UraTyp) developed by the Grambank consortium, which concentrates on broad linguistic variation. The other is the UraTyp questionnaire (UT), designed by the UraTyp authors to represent the internal variation of the Uralic language family. Both questionnaires are organized around structural binary independent variables defining the presence of a feature. The UraTyp webpage revolves around two main interactive tables, Language and Parameters.

The Language table collects genealogical and geographical information on the languages of the sample, including the respective Glottocode and ISO 693-3 code. The Parameters table collects all variables defined by a unique identifier, name, linguistic domain, and a drop-down window that synthetically describes the presence of the variable in the sample. Each variable links to a page describing the feature values and an interactive map. These tables are connected with individual language pages containing the complete set of variables and respective values. In addition, the feature values are plotted on the Uralic phylogenetic tree. The website provides documentation, such as variables discussion and a bibliography. The aggregated data is downloadable in CLDF StructureDataset (Norvik et al. 2022b) for offline use.

## 3.4 Areal genealogically-diverse databases

Areal databases cover genealogically diverse languages spoken in the same broad region. In our survey, these include the Afranaph project, the Hindu Kush areal typology database, the Kiel South Asian typological database, the South American Indigenous Language Structures (SAILS), and WOWA, The Word Order in Western Asia corpus project.

### Afranaph

The Afranaph project (Afranaph; https://afranaphproject.afranaphdatabase.com; Safir 2008) is an online resource that allows for language-specific and cross-linguistic syntactic research in the languages of Africa. The database, originally developed for the study of anaphora, is a living project in constant growth and serves as an umbrella platform for several ongoing and proposed projects, organized through different portals. The Afranaph dataset contains glossed and translated expressions elicited through researcher-prepared questionnaires designed to elicit data that can be used to explore particular areas of grammar. Respondents are consultants with varying amounts of linguistic training who report on their native language. The language sample covers several African genealogically diverse languages, with hundreds of expressions collected for each language. The preparatory materials (from the predefined glossing conventions to the individual language questionnaires) are available to the users for consultation and download.

Although the research behind the development of Afranaph is motivated by generative-grammar-oriented hypotheses, the authors remained theory-neutral in the data presentation, as shown by the architecture of the database, which allows the users to explore the dataset freely, in addition to more theory-constrained queries and properties.

The Afranaph database currently consists of three portals: anaphora, clausal complementation, and generic portal. The generic portal offers an unrestricted exploration of the whole dataset. The portals follow the same general architecture: all include explorative queries, such as search by language or expression, uniquely identified by ISO 693-3 codes and sentence IDs. In addition, users can perform detailed queries by standardized glosses or text content.

The anaphora and the clausal complementation portals offer different analytic entities and search properties. The anaphora portal contains the entity ‘anaphoric marker’ described through complex variables, including morphology, agreement, antecedent properties, locality, predicate compatibility, readings, and pronominal properties. These variables and related sub-variables are described through subsets of fine-grained values (binary and/or categorical).

The clausal complementation portal revolves around analytic entities such as predicate type and meaning properties, c-type and clause type properties, granularly described. The values are either binary or categorical. Query results show relevant glossed expressions. Data and query results have not been exportable for offline use in the past, but as of this writing the project manager reports that this function is in development.

### Hindu Kush areal typology database

The Hindu Kush areal typology database (available through a CLLD web application at https://hindukush.clld.org; Liljegren et al. 2021) is an online resource that collects data on a sample of languages spoken in the Hindu-Kush region. The main goal of the database is to explore contact influence in an area characterized by high typological variation. The languages in the database belong to several distinct linguistic families and subgroups, with data collected from native speakers through questionnaires, wordlists, and video stimuli.

The structural variables in the database are defined binarily (presence vs. absence). The variables describe linguistic information into five macro-domains: clause structure, grammatical categories, lexicon, phonology, and word order. The variables have been chosen based on their representative areality and structural heterogeneity. The database web interface is organized around three primary interactive and downloadable datasets.

The Language dataset contains genealogical and geographic information for each language and unique identifiers such as Glottocodes and ISO 693-3 codes. Users can explore each language separately and visualize its geographical distribution on a map and the list of values for each variable in the dataset. The Feature table lists the name of each variable, its macrodomain, and the respective unique code. Variables have a dedicated page containing a text description of the property, relevant examples, and the feature values for the languages of the sample. The Wordlist table contains interactive maps, IPA transcribed lexical data on the languages of the database, and audio recordings for each entry.
The database is documented internally, through prose descriptions and a detailed bibliography. The aggregated data is downloadable in CLDF StructureDataset (Liljegren et al. 2021) for offline use.

### Kiel South Asian Typological Database

The Kiel South Asian Typological Database (Ivani et al. 2022) is a database focussing on the language variation of several genealogically diverse South Asian languages through hundreds of structural variables. The sample covers three genealogical stocks: Indo-Aryan, Dravidian and Munda, as well as isolates such as Nihali and Kusunda. The data is structured around binary independent variables, uniquely identified, that define the presence or the absence of a given feature. The data comes from descriptive grammars and data collected first-hand during fieldwork trips. The variables list stems from two different questionnaires. One is a subset of the Grambank (see below) questionnaire developed by the Grambank consortium. The other questionnaire targets South Asian specific domains (such as echo constructions) and aims at representing in a fine grained fashion the internal variation of the languages of the Indian subcontinent. The database focuses on morphosyntactic variation, and covers domains such as negation, gender, number, case and non nominative subjects among others. There is no user interface: the data is available as a CSV file freely accessible to the users on the online repository. Basic documentation and variables rationale are available on the online repository.

### South American Indigenous Language Structures online (SAILS online)

The South American Indigenous Language Structures (SAILS; available through a CLLD web application at https://sails.clld.org; Muysken et al. 2016) is a database collecting grammatical properties of more than a hundred languages of South America. A team of linguists has collected data from reference grammars and other descriptive sources. SAILS is composed of several datasets, each with a specific descriptive focus. Some datasets, or domains,[3] limit the scope to a particular geographical area: Arawakan (ARW), Andean (AND), and the Foothills Language (FFQ) domains are among those. Other datasets cover a particular topic: for example, argument marking (ARGEX), subordination (SUB), noun phrase (NP), and tense-aspect-mood and evidentiality (TAME). The linguistic domain-oriented datasets cover roughly the same language sample. SAILS domains code structural properties of languages. The Andean database contains variables sensitive to a specific form or morpheme attested in the language sample. Except for the SUB domain, the variables in the datasets are language-based, where in principle, one language takes a value. The SUB domain data design is construction-based, where constructions take one value from a battery of sub-properties, and languages take multiple constructions. Most variable values throughout SAILS are binary; others are categorical and code phenomena to a set of predefined aggregated types.

[3] We remind the reader that SAILS uses the notion of domain in a broader sense, including geographical areas and language families. In our definition, a domain defines a set of variables that pertain to one particular phenomenon.

SAILS interface revolves around the Languages, Features, and Constructions interactive tables. The Languages table includes data such as ISO 693-3 codes, geographical coordinates, and genealogical information for each language across the domains.

The Features interface allows users to explore at a glance all the variables of the entire SAILS database (except for the SUB ones). Variables are defined by variable name, domain, designer, and number of languages covered for that feature. The Constructions table contains aggregated information on several construction types collected in the SUB domain. In the Construction table, properties are linked to an ID, a construction ID, a description of the property, and the corresponding value.

The SAILS interface allows users to visualize values of interest on the interactive map. The aggregated data is downloadable in CLDF StructureDataset (Muysken et al. 2014) for offline use. Documentation is available on the SAILS webpage and external online repositories (Muysken & O'Connor 2014).

### The Word Order in Western Asia Corpus (WOWA)

The Word Order in Western Asia corpus (WOWA; https://multicast.aspra.uni-bamberg.de/resources/wowa; Haig et al. 2022) is a database containing data on word order on a sample of about 30 genealogically diverse languages of Western Asia. The sample spans eight language families, including Indo-Aryan, Turkic, Indo-Aryan, Iranian, Kartvelian and Semitic.

The aim of the WOWA project is to investigate areal patterns and the impact of language contact on the word order domain. Specifically, the project captures word order patterns on nominal expressions in non-subject position. These are defined by a range of discrete and salient variables (features in WOWA terminology), such as animacy, weight, role and flagging. Each linguistic expression (token) is coded binarily for their position relative to the governing predicate.

The data comes from transcribed spoken corpora, collected within language documentation projects or by accessing primary data or other published resources. Each language (identified by doculect and geographical coordinates) in WOWA has its dataset. This comprises the source text, its syntactically segmented version exported in a spreadsheet template coded for the variables of interest, and metadata. Each dataset includes, on average, half a thousand uniquely identified expressions.

WOWA is documented in detail: the coding guidelines are available to the users in PDF. The source texts are downloadable in PDF and WAV (when available). The spreadsheets with the segmented expressions and the respective coded variables are accessible in CSV and TSV. Metadata is available in PDF.

## 3.5 World-wide databases

Broad cross-linguistic databases cover variation in a maximally diverse sample of languages, usually on a comprehensive battery of linguistic variables. Exceptions include Tymber and the Valency Patterns Leipzig database (ValPal): both are feature-specific databases on a cross-linguistic sample. Unlike most databases (WALS, Terraling, APiCS, Grambank, and AUTOTYP), the Diachronic Atlas of Comparative Linguistics online (DiACL) is enriched by information that is specifically useful for diachronic research.

### Feature-specific typological databases

#### Tymber

Tymber (https://github.com/jkivani/tymber; Ivani & Zakharko 2019) is a typological database on nominal number marking constructions. It contains data on the grammatical number systems of several hundred languages collected from reference grammars and fieldwork materials. Tymber's data consists of fine-grained information on the constructions used to express number categories (singular, dual, plural, including rarer systems such as trial and quadral) on nominals (nouns, pronouns, demonstrative pronouns), semantically defined. Data includes additional properties of the constructions, i.e., the marker type and diachronic information about the etymologies and the proto-forms of the number markers, when available. Tymber relies on the autotypology method (Bickel & Nichols 2002) and the late aggregation principle: the raw data is collected in a bottom-up fashion, stored in a systematic descriptive model, and aggregated only at the analysis phase. This procedure ensures maximal granularity in the description and usability. There is no user interface: the data is distributed over several comma-separated tables in the online data repository. The tables include information on the number categories found on nominals across the sample, organized by language, and the presence of contrastive splits, distinguished by fine-grained semantic properties (i.e., animacy). In addition, the datasets contain aggregated information on the contrastive behavior of groups of nominals in terms of number allomorphy. These groups are defined in Tymber as Reference Types. Additional data consists of information on the construction types (suffixes, reduplication phenomena, and the like) used to mark grammatical number on nominals within and across languages. The languages in the datasets are uniquely identified by Glottocodes and ISO 693-3 codes. The data is downloadable in CSV for offline use.

#### Valency Patterns Leipzig (ValPaL)

The online database Valency Patterns Leipzig (ValPaL; available through a CLLD web application at https://valpal.info, Hartmann et al. 2013) is a large-scale cross-linguistic comparison of valency classes that contains information on several typologically and genealogically diverse languages. ValPal's theoretical approach is inspired by Levin (1993) in applying syntactic diagnostic to identify a semantic classification of verbs. The data in ValPal revolves around the Valency questionnaire. The questionnaire contains a list of predefined verb meanings, purposefully selected for their distinctive cross-linguistic variation in valency behavior and defined by a meaning label, the role frame, and a specific context, such as an example sentence in English. Coding frames contain detailed information on a given verb and its arguments, including their coding and behavioral properties and the relationship of the arguments to the roles in the verb's role frame. Valency alternations define the presence of multiple coding structures associated with members of a set of verb pairs sharing the same verb stem. The domain complexity is encoded by linking multiple simple entities. On the ValPal webpage, these entities are represented by five interactive datasets: Languages, Verb Meanings, Coding Frames, Micro Roles, and Valency Alternations.

The Language dataset assigns each language to the respective Glottocode and provides genealogical and geographical information. The Verb Meaning table illustrates the complete list of verb concepts explored in the database for more than a hundred entries. Each verb meaning is paired with the corresponding Concepticon entry. Concepticon (https://concepticon.clld.org; List et al. 2021) is a broad project that links concept labels from different concept lists to concept sets. The Coding Frames dataset shows the language-specific coding frames identified in the sample. In addition, the coding frame dataset defines whether the coding frame is pre-assessed in the ValPal architecture or derived by alternation.

The other two datasets describe the micro roles and all alternations attested in the database. The Micro Roles dataset contains a label that describes the role ('asker,' 'burdened person'), the corresponding verb meaning, and the semantic role, indicated by the standard linguistic nomenclature (A, P, and the like). The Valency Alternations dataset describes the valency alternation patterns found in the data. The alternations are described through linguistic labels defining the construction ("noun incorporation," "reflexives") and linked to the language they occur in, the respective description, and whether this alternation is coded or uncoded in the dataset. ValPal is documented in detail and the aggregated data is downloadable in CLDF StructureDataset (Forkel 2021) for offline use.

## Broad typological databases

### Diachronic Atlas of Comparative Linguistics Online (DiACL)

The Diachronic Atlas of Comparative Linguistics Online (DiACL; https://diacl.ht.lu.se; Carling 2021) comprises grammatical and lexical data on hundreds of languages from Eurasia, Pacific, and South America. DiACL is primarily designed for comparative and diachronic research. The database contains data on contemporary, historical, and reconstructed languages extracted from descriptive sources and fieldwork materials. The Indo-European family is the linguistic group with the most extensive data coverage and has served as a baseline for developing the database design then applied to other language families.

DiACL core structure revolves around four primary datasets: lexical, typological, metadata, and source data. Lexical data comprises etymological information, Swadesh lists, and a culture vocabulary, organized into semantic classes. Typological/morphosyntactic data includes information on word order, alignment, and nominal/verbal morphology domains. Lexical and typological/morphosyntactic data aim at a "hierarchical organization" (Carling 2021), with the top-most level being the more general. In contrast, lower levels of description are adaptable to areal-specific properties.

The language metadata lists, for each language, a standardized name (and alternative names), ISO 693-3 codes, and Glottocodes. Other information includes geographical location, time frame (an estimation in 100-year intervals within which the language is spoken), language area, and reliability (distinguished in whether the language is modern, dead, and reconstructed). Source data contains reference materials.

The theoretical model that functions as the backbone of the database is a digitized version of the space-time model (Meid 1975), which refers to a stratified model that fixes languages and linguistic patterns in time and space based on historical and contemporary linguistic data (Carling 2021). The primary and unique data points are the languages linked to the respective metadata and phylogenetic information. Languages identifiers are paired with lexical and typological/morphosyntactic data. DiACL datasets can be extracted and downloaded in several formats (CSV, JSON, XLS).

### Syntactic and Semantic Structures of the World's Languages (SSWL, Terraling)

Terraling (https://terraling.com; Koopman and Guardiano 2022) is a collaborative research enterprise hosting several ongoing and proposed linguistic databases. In this contribution, we narrow our discussion to the Syntactic and Semantic Structures of the World's Languages (SSWL) dataset, which has received substantial data coverage.

SSWL contains data on syntactic domains in 300+ languages. SSWL is an open project, regularly updated through the contribution of a broad international team of several hundred linguists, native speakers, and language experts. Documentation on the project is available on an external online repository.

The methodology followed in the data collection procedure aims to answer theoretically guided research in formal syntax and semantics by applying native speakers' introspection on a fine-grained battery of diagnostic tests to sentence properties.

The database is organized around two main datasets: Languages and Properties. The observations are interactive on the webpage, allowing users to jump across datasets and visualize feature-specific values in detail. The Language table assigns each language to the respective ISO 693-3 code and illustrates additional information, such as the percentage of its completeness status in the database and bibliographical references.

The Properties table lists all the syntactic variables covered in SSWL and links each variable to a text description. The description shows examples for each value coded in the database. The values are binary and define the presence or the grammaticality of a specific syntactic pattern. Macro domains, such as word order patterns, are characterized by fine-grained and multiple variables, logically independent from each other. SSWL contains an advanced search engine that compares properties and universal tendencies across languages. Results are exportable in several formats for offline use.

### World Atlas of Language Structures online (WALS online)

The World Atlas of Language Structures online (WALS; available through a CLLD web application at https://wals.info; Dryer et al. 2013) is a feature atlas, published and is maintained by MPI-EVA Leipzig, illustrating the geographical distribution of structural linguistic properties on a worldwide sample. WALS is a joint effort of a broad international team of authors and language experts to provide systematic answers to questions related to the structural diversity and variation in the world's languages. The WALS Online web application supplies a visual overview of this variation through interactive maps, supplemented by detailed feature descriptions. WALS contains more than a hundred chapters, each describing a particular linguistic feature presented in a text and an interactive map. Chapters are grouped thematically in several sections: phonology, morphology, nominal categories, nominal syntax, verbal categories, word order, simple clauses, and complex sentences. WALS contains information on thousands of languages WALS contains information on over 2600 languages overall, but most chapters cover only a few hundred languages. The number and choice of languages vary across the chapters; however, they converge on a 100 (or 200) languages sample, predefined by the WALS editors to maximize consistency and prioritize genealogical and areal diversity across the chapters. Each WALS chapter describes a domain associated with a set of values. Value types vary and range from binary (presence or absence of a specific property) to types to which languages are assigned.
WALS interface consists of several interactive thematic tables. The Features table lists all WALS variables assigned to an ID, the feature name (corresponding to the respective chapter's name), the linguistic domain, and the number of languages covered for that variable. The Chapters table contains the list of chapters in the atlas and provides direct links to each chapter with the variable description. The Languages table allows users to browse the languages in the database, identified by a unique WALS code, the corresponding ISO 693-3 code and the respective Glottocode, and rich genealogical and geographical information. Users can browse data within genealogical groups and restrict the exploration to the 100 or the 200 language samples.
Each feature in the atlas is described by a chapter text and a map. Each chapter provides a detailed description of the feature, the rationale for the respective feature values/types, and practical examples from the sampled languages. Interactive maps illustrate the geographical distribution of the values. Users can combine up to four variables from the atlas through the map tool. WALS aggregated data is downloadable in CLDF StructureDataset (Dryer & Haspelmath 2022) for offline use.

### Atlas of Pidgin and Creole Language Structures online (APiCS online)

The Atlas of Pidgin and Creole Language Structures online (APiCS; available through a CLLD web application at https://apics-online.info; Michaelis et al. 2013a) collects information on about a hundred structural linguistic variables of several pidgins, creoles, and mixed languages worldwide. The data is collected collaboratively by teams of experts, each covering a standardized detailed questionnaire and providing a description survey of a specific language.
APiCS online interface revolves around three primary components, corresponding to individual interactive webpages: Languages, Features, and Survey.
The Languages dataset lists the languages of the sample, plotted on an interactive map, and identified by a progressive number. The table links the languages to several variables: these include a searchable drop-down list of lexifiers, the geographical region, and a link to its Feature dataset. Each Feature dataset contains a short description of the language, its geographical location on a map, and a list of variables with the respective values. Additional data includes downloadable glossed texts, audio recordings and phonological (such as an IPA chart) and sociolinguistic information (number of speakers and level of language endangerment).
Similar to WALS, the variables in APiCS are structural and have fixed values. Unlike WALS, APiCS allows multiple-choice: a language can be coded for more than one type per feature.
The Survey page lists all the surveys included in APiCS. The survey chapters include a language grammatical sketch and a description of its sociolinguistic context.
APiCS contains 18,000 and more linguistic examples. Each example consists of analyzed and glossed texts translated into English. Users can filter through examples by text, gloss, or translation and restrict their search to specific data formats (i.e., audio) or elicitation types.
APiCS variables are connected to WALS parameters through the WALS-APiCS tab. The tab shows the variables shared by both atlases, and users can search through feature names. The aggregated data is downloadable in CLDF StructureDataset (Michaelis et al. 2013b) for offline use.

## Grambank

Grambank (available through a CLLD web application at https://grambank.clld.org; Grambank Consortium 2022) is a worldwide typological database that describes structural linguistic properties in more than 2,600 languages. The project is regularly updated and aims to reach a sample of 3,500 languages. The Grambank database consists of almost two hundred structural variables organized in a questionnaire covering several morphosyntactic domains. An international team of contributors has collected the data through descriptive sources such as reference grammars and consultation with language experts.

While Grambank was inspired by broad typological projects such as WALS, Grambank differs from most WALS descriptions in terms of encoding. WALS partitions phenomena into disjoint categories; domains in Grambank are partitioned along overlapping sets of disjoint categories, and each language is assigned a combination of types from these sets. Grambank feature values are primarily binary, as these define the presence or the absence of a feature. Six feature values, pertaining to word order patterns, are multistate and can easily be binarised.

On the Grambank webpage, users can access genealogical and geographic information about the languages of the sample through the Language page. The Features table collects all the Grambank variables and provides an overview of the distribution of the feature values. Grambank data is downloadable in CLDF StructureDataset for offline use. Extensive documentation and coding guidelines are available on external repositories.

## AUTOTYP

AUTOTYP (https://github.com/autotyp; Bickel et al. 2022) is a large-scale collection of interconnected cross-linguistic databases (modules) with goals in qualitative and quantitative typology (Witzlack et al. 2022). The AUTOTYP project was launched in 1996; since then, it has gone through constant data growth and theoretical and technological developments. AUTOTYP aims to identify structural patterns in the phonological and morphosyntactic domains, assess their genealogical and geographic variation, and ultimately discover the principles governing their distribution (Witzlack et al. 2022). The primary source of the data annotated in AUTOTYP comes from reference grammars and fieldwork materials.

AUTOTYP covers hundreds of typological variables that describe more than a thousand languages over approximately ~260,000 data points (the count includes both primary and derived -aggregated- variables) distributed over several modules.

AUTOTYP relies heavily on several design principles that distinguish it from many traditional typological databases. Firstly, AUTOTYP is modular and interconnected: the various AUTOTYP modules exist as standalone databases and can be connected within AUTOTYP modules and external datasets. In addition, AUTOTYP follows the autotypology method (Bickel and Nichols 2002), which aims to avoid predefined categories in favor of developing them dynamically during data collection. Most typological variables in AUTOTYP are defined in dedicated, continuously re-elaborated definition files and linked to the description of concrete phenomena in data files. The definition files are lists of possible values for each variable coded, and the latter consists of the data on individual languages or constructions in individual languages (Witzlack et al. 2022).

AUTOTYP relies on the principle of late aggregation: the linguistic data is filtered and aggregated during a separate phase through scripts outside the database modules. This procedure ensures data reusability and sustainability (Witzlack et al. 2022). Finally, AUTOTYP implements the exemplar-based method principle: it allows users to select (following algorithmic definitions) exemplars of a particular phenomenon or domains that represent that specific set of values or language.

The *Register* module is a service module. It contains genealogical and geographical information on the languages described in AUTOTYP and other aspects of classification (including Glottocodes and ISO 639-3 codes). The *Register* module includes data on the speakers' primary subsistence (defined binarily whether hunter-gatherer or not), genesis (creole, pidgin, or regular), and modality (spoken or signed language).

The other modules are data modules describing linguistic properties, such as *Categories, Clause, NP, Morphology,* and *GrammaticalRelation*. Each module contains datasets dedicated to individual domains, often compiled with specific research questions and different methodologies combined with several design principles. The description in this subsection is not exhaustive and aims to provide a generic overview of the datasets collected in AUTOTYP.

The module *Categories* comprises datasets describing specific grammatical categories: alienability, clusivity, gender, and numeral classifiers. *Clause* describes selected topics in complex syntax, such as clause linkage and word order. AUTOTYP *Morphology* module contains detailed information on verb morphology and grammatical markers, distributed over five datasets. The module *GrammaticalRelation* provides extensive information on grammatical relations and valence frames. These domains are covered by five datasets, including primary and derived data, the latter processed and aggregated from the primary datasets. Pre-aggregated data from all the AUTOTYP data modules are available in the module *PerLanguageSummaries*, distributed over ~30 datasets.

AUTOTYP data is available in CLDF StructureDataset, CSV and R workspace formats for offline use, while YAML and BIB formats are used for metadata and references.

# 4. Summary

In our survey, we have categorized several properties of linguistic databases and surveyed many of those publicly available.

As a concluding note, and as a desideratum for future modern databases, we would like to encourage database creators to adopt rich descriptive models and prefer fine-grained, high-resolution construction-based approaches for their projects. We firmly believe that the ability to derive aggregated datasets and reuse the data for other research purposes is well worth the additional effort compared with simpler approaches and will greatly enhance the long-term value and usability of databases. Of course, such developments are preconditioned on innovations and more accessible advanced training in linguistic database technology, which will simplify the creation and usage of complex databases. On a related note, it is essential that databases are supplemented with extensive, high-quality documentation as well as systematic metadata to facilitate their usability. We would therefore like to see improvements in the state of the art of database documentation.

# References

Barbiers, Sjef and Leonie Cornips (eds.) 2006. *Dynamic Syntactic Atlas of the Dutch dialects (DynaSAND)*. Amsterdam: Meertens Institut. Available at: http://www.meertens.knaw.nl/sand/. Accessed on: 2022-11-10.

Barbiers, Sjef, Leonie Cornips, and Jan Pieter Kunst. 2007. The Syntactic Atlas of the Dutch Dialects (SAND): A Corpus of Elicited Speech and Text as an Online Dynamic Atlas, in Joan C. Beal, Karen P. Corrigan, Hermann L. Moisl (ed.) *Creating and Digitizing Language Corpora*. London: Palgrave Macmillan. Available at: https://tekstlab.uio.no/nsd. Accessed on: 2022-11-10.

Bickel, Balthasar and Johanna Nichols. 2002. Autotypologizing databases and their use in fieldwork, in Peter, Austin, Helen Dry, Peter Wittenburg (ed.) *Proceedings of the International LREC Workshop on Resources and Tools in Field Linguistics*. Las Palmas: Nijmegen: MPI for Psycholinguistics.

Bickel, Balthasar, Johanna Nichols, Taras Zakharko, Alena Witzlack-Makarevich, Kristine Hildebrandt, Michael Rießler, Lennart Bierkandt, Fernando Zúñiga & John B Lowe. 2022. The AUTOTYP database (v1.0.1). Zenodo. https://doi.org/10.5281/zenodo.6255206. Accessed on: 2022-11-10.

Carling, Gerd (ed.) 2017. *Diachronic Atlas of Comparative Linguistics Online*. Lund: Lund University. Available at: https://diacl.ht.lu.se/. Accessed on: 2022-11-10.

Carling, Gerd (ed.) 2021. *Diachronic Atlas of Comparative Linguistics Online: Database description (2.1)*. Lund: Lund University, Centre for Languages and Literature.

Corbett, Greville G. 2006. *Agreement*. Cambridge: Cambridge University Press.

Cristofaro, Sonia. 2013. Purpose Clauses. In: Dryer, Matthew S. & Haspelmath, Martin (eds.) The World Atlas of Language Structures Online. Leipzig: Max Planck Institute for Evolutionary Anthropology. (Available online at http://wals.info/chapter/125). Accessed on: 2022-11-10.

Cysouw, Michael, Jeff Good, Mihai Albu & Hans-Jörg Bibiko. 2005. Can GOLD “cope” with WALS? Retrofitting an ontology onto the World Atlas of Language Structures. In *Proceedings of E-MELD 2005: Linguistic ontologies and data categories for language resources*, Cambridge, Mass., July 1–3.

Dryer, Matthew S and Martin Haspelmath. (eds). 2013. *WALS Online*. Leipzig: Max Planck Institute for Evolutionary Anthropology. Available at: https://wals.info/. Accessed on: 2022-11-10.

Dryer Matthew, and Martin Haspelmath. 2022. The World Atlas of Language Structures Online (v2020.2) [Data set]. Zenodo. https://doi.org/10.5281/zenodo.6806407. Accessed on: 2022-11-10.

Farrar, Scott. and Langendoen, D.Terence. 2003. A linguistic ontology for the semantic web. *GLOT international*, *7*(3), pp.97-100.

Fernández, Beatriz, Ane Berro, Iñigo Urrestarazu, and Itziar Orbegozo. 2019. Mapping variation in Basque: The BIV database, *Linguistic Typology*, 23(2), pp. 347–374.

Grambank Consortium (eds.) 2022. *Grambank*. Leipzig: Max Planck Institute for Evolutionary Anthropology. Available at: http://grambank.clld.org/. Accessed on: 2022-11-10.

Forkel, Robert, Johann-Mattis List, Simon J. Greenhill, Christoph Rzymski, Sebastian Bank, Michael Cysouw, Harald Hammarström, Martin Haspelmath, Gereon A. Kaiping, and Russell D. Gray. 2018. Cross-Linguistic Data Formats, advancing data sharing and re-use in comparative linguistics. *Scientific Data* 5, no. 1: 1-10.

Forkel, Robert, Sebastian Bank, Christoph Rzymski, and Hans-Jörg Bibiko. 2020. clld/clld: clld - a toolkit for cross-linguistic databases (v7.2.0). Zenodo. https://doi.org/10.5281/zenodo.3968247. Accessed on: 2022-11-10.

Forkel, Robert. 2021. CLDF dataset derived from Hartmann et al.'s "Valency Patterns Leipzig" from 2013 (v1.0) [Data set]. Zenodo. https://doi.org/10.5281/zenodo.5667260

Hammarström, Harald, Robert Forkel, Martin Haspelmath, and Sebastian Bank (eds.) 2021. *Glottolog 4.5*. Leipzig: Max Planck Institute for Evolutionary Anthropology. Available at: http://glottolog.org/. Accessed on: 2022-11-10.

Haig, Geoffrey & Stilo, Donald & Doğan, Mahîr C. & Schiborr, Nils N. (eds.). 2022. *WOWA — Word Order in Western Asia: A spoken-language-based corpus for investigating areal effects in word order variation*. Bamberg: University of Bamberg (multicast.aspra.uni-bamberg.de/resources/wowa/). Accessed on: 2022-11-10.

Hartmann, Iren, Martin Haspelmath, and Bradley Taylor (eds.) 2013. *The Valency Patterns Leipzig online database*. Leipzig: Max Planck Institute for Evolutionary Anthropology. Available at: https://valpal.info/. Accessed on: 2022-11-10.

Idone, Alice. 2018. 'Il progetto The Zurich database of agreement in Italo-Romance', in Gianna Marcato (ed.) *Dialetto e società. Presentazione di lavori in corso*, pp. 15–22.

Ivani, Jessica K. and Taras Zakharko. 2019. *Tymber, Database on nominal number marking constructions*. Dataset. Available at: https://github.com/jivani/tymber/. Accessed on: 2022-11-10.

Ivani, Jessica K, Peterson, John & Chevallier, Lennart. 2022. *The Kiel South Asian Typological Database.* Dataset. [https://doi.org/10.5281/zenodo.7153825]. Accessed on: 2022-11-10.

Johannessen, Janne Bondi, Joel Priestley, Kristin Hagen, Tor Anders Åfarli and Øystein A. Vangsnes (2009) 'The Nordic Dialect Corpus - an Advanced Research Tool', in Kristiina Jokinen and Eckhard Bick. (eds.) *Proceedings of the 17th Nordic Conference of Computational Linguistics NODALIDA 2009. NEALT Proceedings Series Volume 4.*, pp. 283–286. Available at: https://tekstlab.uio.no/nsd.

Koopman, Hilda, and Cristina Guardiano. 2022. Managing Data in TerraLing, a Large-Scale Cross-Linguistic Database of Morphological, Syntactic, and Semantic Pattern, in Andrea L.

Berez-Kroeker, L.B.C., Bradley McDonnell, Eve Koller (eds.) *The Open Handbook of Linguistic Data Management*. The MIT Press.

Kortmann, Bernd, Kerstin Lunkenheimer, and Katharina Ehret (eds.) 2020a. *eWAVE*. Available at: https://ewave-atlas.org/. Accessed on: 2022-11-10.

Kortmann, Bernd, Kerstin Lunkenheimer, & Katharina Ehret. 2020b. cldf-datasets/ewave: The Electronic World Atlas of Varieties of English (v3.0) [Data set]. Zenodo. https://doi.org/10.5281/zenodo.3712132. Accessed on: 2022-11-10.

Levin, Beth. 1993. *English Verb Classes and Alternations*. Chicago: University of Chicago Press.

Liljegren, Henrik, Robert Forkel, Nina Knobloch, and Noa Lange (ed.) 2021. *Hindu Kush Areal Typology (Version v1.0)*. Dataset. Zenodo. http://doi.org/10.5281/zenodo.4534221. Accessed on: 2022-11-10.

Lindstad, Arne Martinus, Anders Nøklestad, Janne Bondi Johannessen, and Øystein Alexander Vangsnes. 2009. The Nordic Dialect Database: Mapping Microsyntactic Variation in the Scandinavian Languages. In *Proceedings of the 17th Nordic Conference of Computational Linguistics (NODALIDA 2009)*, 283–286, Odense, Denmark. Northern European Association for Language Technology (NEALT).

List, Johann M., Annika Tjuka, Christoph Rzymski, Simon Greenhill, Nathanael Schweikhard, and Robert Forkel (eds.) 2021. *Concepticon 2.5.0*. Leipzig: Max Planck Institute for Evolutionary Anthropology. Available at: https://concepticon.clld.org/. Accessed on: 2022-11-10.

Longobardi, Giuseppe. 2003. *Methods in Parametric Linguistics and Cognitive History*. Linguistic Variation Yearbook 3(1). 101–138.

Loporcaro, Michele, Taras Zakharko, Tania Paciaroni, Diego Pescarini, Alice Idone, Serena Romagnoli, and Chiara Zanini (eds.) 2018. *The Zurich Database of Agreement in Italo-Romance (DAI)*. Zürich: University of Zürich. Available at: https://www.dai.uzh.ch/. Accessed on: 2022-11-10.

Meid, Wolfgang. 1975 *Probleme der räumlichen und zeitlichen Gliederung des Indogermanischen*. Reichert.

Michaelis, Susanne M, Philippe Maurer, Martin Haspelmath, and Magnus Huber (eds.) 2013a. *Atlas of Pidgin and Creole Language Structures Online*. Leipzig: Max Planck Institute for Evolutionary Anthropology. Available at: https://apics-online.info/. Accessed on: 2022-11-10.

Michaelis, Susanne M, Philippe Maurer, Martin Haspelmath, & Magnus Huber. 2013b. cldf-datasets/apics: The "Atlas of Pidgin and Creole Language Structures Online" as CLDF dataset (Version v2013) [Data set]. Zenodo. https://doi.org/10.5281/zenodo.3823888. Accessed on: 2022-11-10.

Muysken, Pieter, Harald Hammarström, Olga Krasnoukhova, Neele Müller, Joshua Birchall, Simon van de Kerke, Loretta O'Connor, Swintha Danielsen, Rik van Gijn, & George Saad. 2014. cldf-datasets/sails: South American Indigenous Language Structures (SAILS) (Version v2014) [Data set]. Zenodo. https://doi.org/10.5281/zenodo.3608862. Accessed on: 2022-11-10.

Muysken, Pieter & Loretta O'Connor (eds.). 2014. The Native Languages of South America: Origins, Development, Typology. Cambridge: Cambridge University Press.

Muysken, Pieter, Harald Hammarström, Olga Krasnoukhova, Neele Müller, Joshua Birchall, Simon van de Kerke, Loretta O'Connor, Swintha Danielsen, Rik van Gijn, and George Saad (eds.) 2016. *South American Indigenous Language Structures (SAILS) Online*. Jena: Max Planck Institute for the Science of Human History. Available at: https://sails.clld.org. Accessed on: 2022-11-10.

Norvik, Miina, Sirkka Saarinen, Yingqi Jing, Michael Dunn, Robert Forkel, Terhi Honkola, Gerson Klumpp, Richard Kowalik, Helle Metslang, Karl Pajusalu, Minerva Piha, Eva Saar, and Outi Vesakoski. 2022a. 'Uralic Typology in the light of new comprehensive data sets'. In *Journal of Uralic Linguistics, Volume 1, Issue 1.*

Norvik, Miina, Yingqi Jing, Michael Dunn, Robert Forkel, Terhi Honkola, Gerson Klumpp, Richard Kowalik, Helle Metslang, Karl Pajusalu, Minerva Piha, Eva Saar, Sirkka Saarinen, & Outi Vesakoski. 2022b. Uralic Typological database - UraTyp (v1.1) [Data set]. Zenodo. https://doi.org/10.5281/zenodo.6392555. Accessed on: 2022-11-10.

Orbegozo, Itziar, Iñigo Urrestarazu, Ane Berro, Josu Landa, and Beatriz Fernández. 2018. *Euskara Bariazioan / Basque in Variation (BiV) (2nd ed.)*. UPV/EHU. Available at: https://basqueandbeyond.ehu.eus/Bas&Be/biv/. Accessed on: 2022-11-10.

Poletto, Chiara, and Paola Benincà. 2007. 'The ASIS enterprise: a view on the construction of a syntactic atlas for the Northern Italian dialects', *Nordlyd : Tromsø University Working Papers on Language & Linguistics / Institutt for Språk og Litteratur, Universitetet i Tromsø*, 34. doi:10.7557/12.88.

Safir, Ken (ed.) 2008. *Afranaph Database*. Available at https://www.africananaphora.rutgers.edu. Accessed on: 2022-11-10.

Seiler, Guido, Sandro Bachmann, Johannes Graën, Nikolina Rajović, Adrian van der Lek, Ghazi Hachfi, Igor Mustač, Elvira Glaser, Peter Ranacher, Robert Weibel. 2022. *Syntaktischer Atlas der deutschen Schweiz online (SADS online)*. Deutsches Seminar / Linguistic Research Infrastructure / UFSP Sprache und Raum, Universität Zürich.

Stassen, Leon. 2013. Comparative Constructions. In: Dryer, Matthew S. & Haspelmath, Martin (eds.) The World Atlas of Language Structures Online. Leipzig: Max Planck Institute for Evolutionary Anthropology. (Available online at http://wals.info/chapter/121). Accessed on: 2022-11-10.

Witzlack-Makarevich, Alena, Johanna Nichols, Kristine A. Hildebrandt, Taras Zakharko, and Balthasar Bickel. 2022. Managing AUTOTYP Data: Design Principles and Implementation, Andrea L. Berez-Kroeker, L.B.C., Bradley McDonnell, Eve Koller (eds.) *The Open Handbook of Linguistic Data Management*. The MIT Press.

# Appendix

| Database | URL | Topics | Areality | Design principle | Variables | Unit of description | Documentation | Query interface | Query interface usability | Raw data export | Language reports | Interoperability |
|---|---|---|---|---|---|---|---|---|---|---|---|---|
| BiV | https://basqueandbeyond.ehu.eus/biv/ | Case and agreement; auxiliary alternation; aspect; postpositions; transitivity; complementizers; mood | Basque varieties | Monocategorization | Presence/absence of standard patterns vs variational patterns | Language | Limited | Cartography, features chapters | Easy | CSV, XLS, XML | Yes | No |
| DynaSAND | https://www.meertens.knaw.nl/sand/ | Auxiliaries and verb position; complementizers; expletive subjects; form of subject pronouns; verbal clusters; negation and quantification; one pronominalization; reflexive and reciprocal pronouns; relative clauses; | Dutch varieties | Monocategorization | Elicited written and oral questionnaires tagged by syntactic domain, sociolinguistic properties and reliability judgments | Expression | Limited | Corpus search, cartography | Involved | CSV, XLS | Yes | Partial (Kloeke-codes) |
| | | subject doubling; subject clitics | | | | | | | | | | |
| SADS | https://dialektsyntax.linguistik.uzh.ch | NP, VP, pronouns, simple and complex syntax | Swiss German varieties | Monocategorization | Presence/absence of standard patterns vs variational patterns | Language | Limited | Cartography, simple queries | Easy | CSV | Yes | No |
| eWAVE | https://ewave-atlas.org | Pronouns; NP; TAM; verb morphology; negation; agreement; relativization; complementation; adverbial subordination; adverbs and prepositions; discourse and word order | English varieties | Monocategorization | Presence/absence and pervasiveness of predefined categories | Language | Extensive | Cartography, filter search, drop-down search, features chapters | Easy | CLDF | No | ISO 693-3 codes, Glottocodes |
| ASIt | http://asit.maldura.unipd.it | Adverbs; clitics; negation; object; quantification; NP; PP; relative clauses; subject; verb; interrogative sentences; exclamative sentences | Languages of Italy | Monocategorization | Elicited sentences tagged by syntactic keywords | Expression | None | Syntactic search | Difficult | None | Yes | No |
| ScanDyaSYN | http://websim.arkivert.uit.no/scandiasyn/?Language=en | Binding and co-reference; left and right periphery; subject placement; object shift; auxiliary, tense marking; NP; VP; argument structure; verb placement | Danish, Faroese, Icelandic, Norwegian, Swedish | Monocategorization | Transcribed speech data tagged by syntactic domain and sociolinguistic properties | Expression | Limited | Corpus search, sociolinguistic, cartography | Involved | None | Yes | No |
| DAI | https://www.dai.uzh.ch/new/#/public/home | Agreement | Italo-Romance | Structural decomposition | Elicited sentences and speech data tagged by PoS, lemmas, morphosyntactic patterns | Expression | Limited | PoS and lemmas search, agreement patterns search, sociolinguistic | Involved | None | Yes | No |
| UraTyp | https://uralic.clld.org | Broad morphosyntactic features | Uralic | Monocategorization, Multicategorization | Presence/absence of various typological structural and | Language | Extensive | Cartography, filter search, drop- | Easy | CLDF | No | ISO 693-3 codes, Glottocodes |
| | | | | | genealogical features | | | down search | | | | |
| Afranaph | https://afranaphproject.afranaphdatabase.com | Anaphora; clausal complementation | Languages of Africa | Multicategorization | Elicited written sentences in a standardized questionnaire tagged by syntactic categories, glosses and text | Expression | Extensive | Filter search, text search, syntactic search | Difficult | None | Yes | ISO 693-3 |
| HinduKush | https://hindukush.clld.org | Clause structure; grammatical categories; lexicon, phonology; word order | Languages of the Hindu-Kush region | Monocategorization | Presence/absence of various simple structural and areal features | Language | Extensive | Cartography, filter search, drop-down search | Easy | CLDF | No | ISO 693-3 codes, Glottocodes |
| Kiel South Asian Typological Database | https://zenodo.org/record/7153825 | Grammatical categories; morphosyntax, negation | Indo-Aryan, Dravidian, Munda, isolate of South Asia | Monocategorization, Multicategorization | Presence/absence of various typological structural and genealogical features | Language | Minimal | External | NA | CSV | No | Glottocodes |
| SAILS | https://sails.clld.org | Argument marking, subordination, NP, TAME, areal specific features | Languages of South America | Monocategorization, Multicategorization | Presence/absence of structural and areal features, construction-based description | Language; Construction | Extensive | Cartography, filter search, drop-down search | Easy | CLDF | No | IISO 693-3 codes, Glottocodes |
| WOWA | https://multicast.aspra.uni-bamberg.de/resources/wowa/ | Word order of nominal expressions | Languages of Western Asia | Monocategorization | Presence/absence of structural features, position | Expression | Extensive | External | NA | CSV, TSV | Yes | No |
| Tymber | https://github.com/jkivani/tymber | Grammatical number marking (nouns, pronouns, demonstrative pronouns) | World-wide | Structural decomposition | Presence/absence of structural properties, construction-based, autotypology | Construction; Language | Minimal | External | NA | CSV | No | ISO 693-3 codes, Glottocodes |
| ValPal | https://valpal.info | Valency | World-wide | Structural decomposition | Valency frames, predicate behavior, verb meanings, microroles alternations defined by a standardized questionnaire | Construction | Extensive | Cartography, filter search, drop-down search | Easy | CLDF | Yes | IISO 693-3 codes, Glottocodes |
| DiACL | https://diacl.ht.lu.se | Broad linguistic domains, lexicon | World-wide | Monocategorization | Presence/absence of typological structural features and lexical data for comparative and diachronic research | Language | Limited | Carthography, external | Involved | CSV (partial), JSON, XLS | No | ISO 693-3 codes, Glottocodes |
| SSWL | https://terraling.com | Word order | World-wide | Monocategorization | Presence/absence of structural features | Language | Minimal | Simple and combined queries | Involved | JSON | Partial | ISO 693-3 codes, Glottocodes |
| WALS | https://wals.info | Broad linguistic domains | World-wide | Monocategorization | Presence/absence of typological structural features, predefined types | Language | Extensive | Carthography, combine, filter, drop-down search | Easy | CLDF | No | IISO 693-3 codes, Glottocodes |
| APiCS | https://apics-online.info | Broad linguistic domains | Pidgin and creoles | Monocategorization | Presence/absence of typological structural | Language | Extensive | Carthography, combine, filter, | Easy | CLDF | Yes | ISO 693-3 codes, Glottocodes |
| | | | | | features, predefined types | | | drop-down search | | | | |
| GramBank | https://grambank.clld.org | Broad morphosyntactic features | World-wide | Multicategorization | Presence/absence of typological structural features | Language | Extensive | Cartography, filter search, drop-down search | Easy | CLDF | No | Glottocodes |
| AUTOTYP | https://github.com/autotyp | Grammatical categories; grammatical relations; clause linkage; clause order; morphology; phonological properties | World-wide | Monocategorization, multicategorization, structural decomposition | Presence/absence of structural properties, construction-based, autotypology | Construction; Language | Limited | External | NA | CSV, JSON, CLDF, R | Partial | ISO 693-3 codes, Glottocodes |